\documentclass[twocolumn]{aastex61}

\usepackage{natbib}

\usepackage{afterpage}
\usepackage{tensor}
\usepackage{graphicx}
\usepackage{subfigure}
\usepackage{amsmath, amssymb, amsfonts, mathtools}
\usepackage{ulem}

\newcommand \Msun {\ensuremath{ \, \rm M_{\sun} }}  
\newcommand \msun { \Msun }

\renewcommand{\vec}[1]{\boldsymbol{#1}} 

\newcommand \rmag {\ensuremath{ R_{\rm m}  }}
\newcommand \rco  {\ensuremath{ R_{\rm co} }}
\newcommand \rlc  {\ensuremath{ R_{\rm LC} }}

\newcommand \G {\, {\rm G}}

\newcommand \ergsec {\, \rm{erg \, s^{-1}}}

\newcommand \km {\, {\rm km}}
\newcommand \Hz {\, {\rm Hz}}

\newcommand \rg     {\ensuremath{ r_{\rm g}      }}
\newcommand \rgu    {\ensuremath{ \, r_{\rm g}      }}
\newcommand \rgc    {\ensuremath{ \, r_{\rm g}/c }}

\newcommand \ffrac {\mathcal{F}}
\newcommand \profile {\mathcal{P}}
\newcommand \smooth {\mathcal{A}}
\newcommand \uint {\epsilon}
\newcommand \muu {\mu_{\rm unit}}

\shorttitle{Accretion onto Magnetized, Rotating Neutron Stars}
\shortauthors{Parfrey \& Tchekhovskoy}

\begin{document}

\title{General-Relativistic Simulations of Four States of Accretion onto Millisecond Pulsars}

\author{Kyle Parfrey}
\altaffiliation{Einstein Fellow}
\affiliation{Lawrence Berkeley National Laboratory, 1 Cyclotron Road, Berkeley, CA 94720}
\affiliation{Departments of Physics and Astronomy, and Theoretical Astrophysics Center, UC Berkeley, CA 94720}

\author{Alexander Tchekhovskoy}
\altaffiliation{Theoretical Astrophysics Center Fellow}
\affiliation{Departments of Physics and Astronomy, and Theoretical Astrophysics Center, UC Berkeley, CA 94720}
\affiliation{Center for Interdisciplinary Exploration \& Research in Astrophysics (CIERA),
Physics \& Astronomy, Northwestern University, Evanston, IL 60202}
\affiliation{Lawrence Berkeley National Laboratory, 1 Cyclotron Road, Berkeley, CA 94720}
\affiliation{Kavli Institute for Theoretical Physics, Kohn Hall, University
of California at Santa Barbara, Santa Barbara, CA 93106}

\email{kparfrey@lbl.gov}

\begin{abstract} 
Accreting neutron stars can power a wide range of astrophysical
phenomena including short- and long-duration gamma-ray bursts,
ultra-luminous X-ray sources, and X-ray binaries. 
Numerical simulations are a valuable tool for
studying the accretion-disk--magnetosphere interaction that is central
to these problems, most clearly for the recently discovered transitional millisecond pulsars.
However, magnetohydrodynamic (MHD) methods, widely used for simulating accretion, have
difficulty in highly magnetized stellar magnetospheres, 
while force-free methods, suitable for such regions, cannot include the accreting gas.
We present an MHD method that can stably evolve essentially force-free,
highly magnetized regions, and describe the first time-dependent
relativistic simulations of magnetized accretion onto millisecond
pulsars.  Our axisymmetric general-relativistic MHD simulations for
the first time demonstrate how the interaction of a turbulent
accretion flow with a pulsar's electromagnetic wind can lead to the transition
of an isolated pulsar to the accreting state. This transition naturally
leads to the formation of relativistic jets, whose power can greatly
exceed the power of the isolated pulsar's wind. If the accretion rate is
below a critical value, the pulsar instead expels the accretion stream. 
More generally, our simulations
produce for the first time the four possible accretion regimes, in
order of decreasing mass accretion rate: (a) crushed magnetosphere
and direct accretion; (b) magnetically channeled accretion onto the
stellar poles; (c) the propeller state, where material enters
through the light cylinder but is prevented from accreting by the
centrifugal barrier; (d) almost perfect exclusion of the accretion
flow from the light cylinder by the pulsar wind.
\end{abstract}


\section{Introduction} 
Accretion onto rotating neutron stars is the principal agent
underlying several classes of high-energy objects, including nuclear-
and accretion-powered millisecond X-ray pulsars
\citep[AMXPs;][]{van-der-Klis:2000,Patruno:2012aa}, the atoll sources and the
rapidly accreting Z sources \citep{Hasinger:1989aa}. Two particularly
interesting varieties were recently discovered: the transitional
millisecond pulsars, which switch between X-ray and radio
pulsing states \citep{Papitto:2013aa,Stappers:2014aa}, and the highly
super-Eddington pulsing ultra-luminous X-ray sources
\citep[ULXs;][]{Bachetti:2014aa,Furst:2016aa,Israel:2017ab}. Making
sense of these objects requires understanding how 
turbulent, magnetized accretion flows interact with 
stellar magnetospheres.

Non-relativistic MHD simulations have been extensively employed to
study accretion onto magnetized stars, displaying plasmoid ejections,
accretion columns, magnetically powered jets, and
centrifugal inhibition of accretion
\citep[e.g.][]{Hayashi:1996aa,Miller:1997aa,Romanova:2002aa,Kato:2004aa,Zanni:2009aa,Romanova:2012aa,Lii:2014aa}. 
Simulations of this kind become more challenging in the relativistic regime,
because of the need to maintain nearly force-free (highly magnetically
dominated) regions alongside the accreting plasma. Relativistic MHD
simulations of isolated-pulsar magnetospheres have been
performed by modifying the evolution of the densities and velocities
\citep*{Komissarov:2006aa,Tchekhovskoy:2013aa}. On the other hand,
force-free electrodynamics, a simplified form of high-magnetization
MHD, has been used to study the interaction of pulsar magnetospheres
with accretion disks, which were kinematically specified rather than dynamically evolved
\citep*{Parfrey:2017aa}. Here we describe a new approach to seamlessly
combine in one simulation a full-MHD accretion disk 
interacting with an essentially force-free stellar magnetosphere.

\section{Numerical approach}

We carry out the simulations in spherical coordinates ($r,\theta,\phi$) with a modified 
version of the \textsc{harmpi} code\footnote{https://github.com/atchekho/harmpi}, 
which is based on \textsc{harm2d} \citep{Gammie:2003aa,Noble:2006aa}. 
We evolve the ideal GRMHD equations,
\begin{align}
  \nabla_\mu (\rho u^\mu)&=0,\nonumber\\
  \nabla_\mu T^{\mu}_{\nu}&=0,\label{eq:mhd}\\
  \nabla_\mu F^{*\mu\nu}&=0,\nonumber
\end{align}
where 
\begin{equation}
\label{eq:tmunu}
  T^{\mu\nu}=\left(\rho+\frac{\uint+p}{c^2}+\frac{b^2}{4\pi c^2}\right)u^\mu u^\nu+\left(p+\frac{b^2}{8\pi}\right)g^{\mu\nu}-\frac{b^\mu b^\nu}{4\pi}
\end{equation}
is the total energy-momentum tensor; $\rho$, $\uint$, and $p$ are the 
fluid-frame
mass density, internal-energy density, and gas pressure; $u^\mu$ is the fluid four-velocity; $b^\mu$ is the
magnetic four-vector ($b^2/8\pi$ is the fluid-frame magnetic pressure);
and $F^{*\mu\nu}=b^{\mu}u^{\nu}-b^{\nu}u^{\mu}$ is the dual of the electromagnetic field tensor.
Vector components written as $V_{\hat{\alpha}}$ have the same dimensions as the physical quantity
they represent, and are defined as $V_{\hat{\phi}}=\mathrm{sign}(V^{\phi})\sqrt{V_{\phi}V^\phi}$.

After the solution is stepped forward in time according to Eqn.~(\ref{eq:mhd}), we adjust 
the hydrodynamic variables $\rho$, $\uint$, and $u^\mu$ inside the force-free magnetosphere for 
improved behavior in the high-magnetization regime.

To distinguish the force-free magnetosphere from the accretion flow,
we introduce a new variable $\ffrac$ which is evolved as a passive
scalar, $\nabla_\mu(\ffrac\rho u^\mu)=0$. $\ffrac$ is the
fraction of a given zone's mass density that can be identified as
magnetospheric, or due to the density ``floor'' which is required for
numerical stability: $\ffrac=1$ in the isolated-pulsar magnetosphere
before the onset of accretion, and $\ffrac\approx0$ inside the
accretion flow. Additionally, we specify a smooth, fixed radial profile
$\profile(r)$, whose purpose is to decrease the severity of the adjustments with increasing
$r$, such that they vanish beyond the light cylinder: $\profile=1$ at the stellar surface, $r=r_*$, and
$\profile=0$ for $r\geq\rlc$, where $\rlc=c/\Omega$ is the light-cylinder (LC) radius
and $\Omega$ is the stellar spin angular frequency.
We make all adjustments according to the value of $\smooth=1-\ffrac\profile$.

When $\smooth=1$ (i.e., inside the accretion flow or in any cell beyond the light cylinder) 
the evolution is unmodified and follows the GRMHD equations. 
When $\smooth=0$ the mass and internal-energy densities are set to fixed background
distributions,
and the velocity parallel to the magnetic field, as measured by the 
coordinate-static observer,\footnote{Coordinate-static observer: $n^\alpha=(n^t,0^i)$} is set to zero. 
For $0<\smooth<1$ the modification to these quantities interpolates linearly between
these two extremes. When mass or internal energy is added to (or removed from) a cell, the fluid 
velocity is adjusted, such that the total conserved momentum parallel to the magnetic field, as
measured by the coordinate-static observer, is unchanged.

At the inner radial boundary, $r=r_*$, we extrapolate the mass and internal-energy
densities into the star. We fix the radial magnetic field 
to its dipolar distribution, and extrapolate the tangential components.
For surface locations at which accretion occurs, we set the velocities to corotation with the 
prescribed stellar angular velocity $\Omega$, plus the extrapolated value of the velocity along
the magnetic field, as measured by the corotating observer.\footnote{Corotating observer: $n^\alpha=n^t(1,0,0,\Omega)$} 
For non-accreting locations we set the velocity 
such that the coordinate-static observer measures it to be the component of the stellar angular 
velocity that is perpendicular to the magnetic field.

We have confirmed that in flat spacetime, using our boundary conditions and 
treatment of magnetically dominated regions, we reproduce the analytic solution for a rotating 
monopole \citep{Michel:1973ab} and the flux distribution and spin-down power of a rotating 
dipole \citep{Contopoulos:1999aa,Gruzinov:2005aa}. The numerical methods and associated tests will be
described in detail in a future paper.

\section{Problem Configuration}

We initialize the problem with an 
equilibrium hydrodynamic torus \citep{Chakrabarti:1985,De-Villiers:2003aa}, with the inner edge at $r=60\rgu$, where 
$\rg=GM/c^2$ is the gravitational radius, density maximum $\rho_{\rm max}=1$ at 
$r_{\rm max}=85\rgu$, and angular momentum distribution $\propto r^{1/3}$. 
We insert a magnetic field loop into the torus, with plasma $\beta=8\pi p/b^2\ge100$ and poloidal field lines 
following density isosurfaces, $A_\phi\propto\rho$, and add random, 
$10\%$-level pressure variations to jump-start the magnetorotational instability (MRI; \citealt{Balbus:1991aa}). 
We consider two orientations of the torus magnetic field:  
parallel or anti-parallel to the stellar dipole field at the equator.

The magnetosphere, occupying the entire computational domain outside the torus, is dominated by the 
neutron star's magnetic field. We initialize it using the vector potential of a static, potential dipole of 
magnetic moment $\mu$ in the Schwarzschild spacetime \citep{Wasserman:1983aa},
measured in units of $\muu=\rg^3\sqrt{4\pi\rho_{\rm max}c^2}$.
For rotating stars, we deform the stellar field lines, which would be rotationally opened by an isolated pulsar, 
to go around and close beyond the torus. This prevents these field lines from becoming artificially 
trapped by the torus when the star begins to rotate.

Near the star the gas and internal-energy densities create a 
hydrostatic atmosphere, with zero velocity relative to the coordinates, that smoothly transitions 
to $\rho,\epsilon\propto r^{-6}$ and $\epsilon/\rho=0.035$ at $r\gtrsim\rlc$. 
The magnetosphere is highly magnetically dominated, with the initial magnetization near the star 
$\sigma=b^2/4\pi\rho{}c^2\sim3\times10^4\gg1$. 

We performed 19 simulations in two sets. In the first, $\mu$ = [$1.25,2.5,5,10,20,40,80,160$],
$\rlc=20\rgu$, for both orientations of the torus magnetic field. In the second, $\mu=20$, 
$\rlc/\rg$ = [$20,40,60,\infty$], for the anti-parallel stellar-torus 
field orientation. 
We ran the simulations for $\Delta t=5\times10^4\rgc$, or $\sim10$ Keplerian orbits at $r_{\rm max}$.

We set the stellar radius to $r_*=4\rgu$; this choice for typical
$r_*=10\km$ gives the spin frequency $\nu\approx950(\rlc/20\rgu)^{-1}\Hz$. We use the Kerr 
spacetime in the Boyer-Lindquist foliation, with spin parameter $a=(1/3)(\rlc/20\rgu)^{-1}$
\citep[e.g.][]{Belyaev:2016aa}; the results are essentially unchanged when we instead use the Kerr-Schild
spacetime foliation.

The numerical resolution is $N_r\times{}N_\theta\times{}N_\varphi=1024\times512\times1$ in the domain 
extending to $r_{\rm out}=5\times10^3\rgu$. 
The radial grid is logarithmically spaced for $r\le200\rgu$, beyond which it becomes 
hyper-exponential \citep{Tchekhovskoy:2009aa}; 
the angular grid is mildly concentrated toward the midplane. 
We use Lax-Friedrichs fluxes with the monotonized-central limiter.

\section{Results}

\begin{figure*}
  \begin{center}
    \includegraphics[width=\textwidth, trim = 2.0mm 2.0mm 2.0mm 2.0mm, clip]{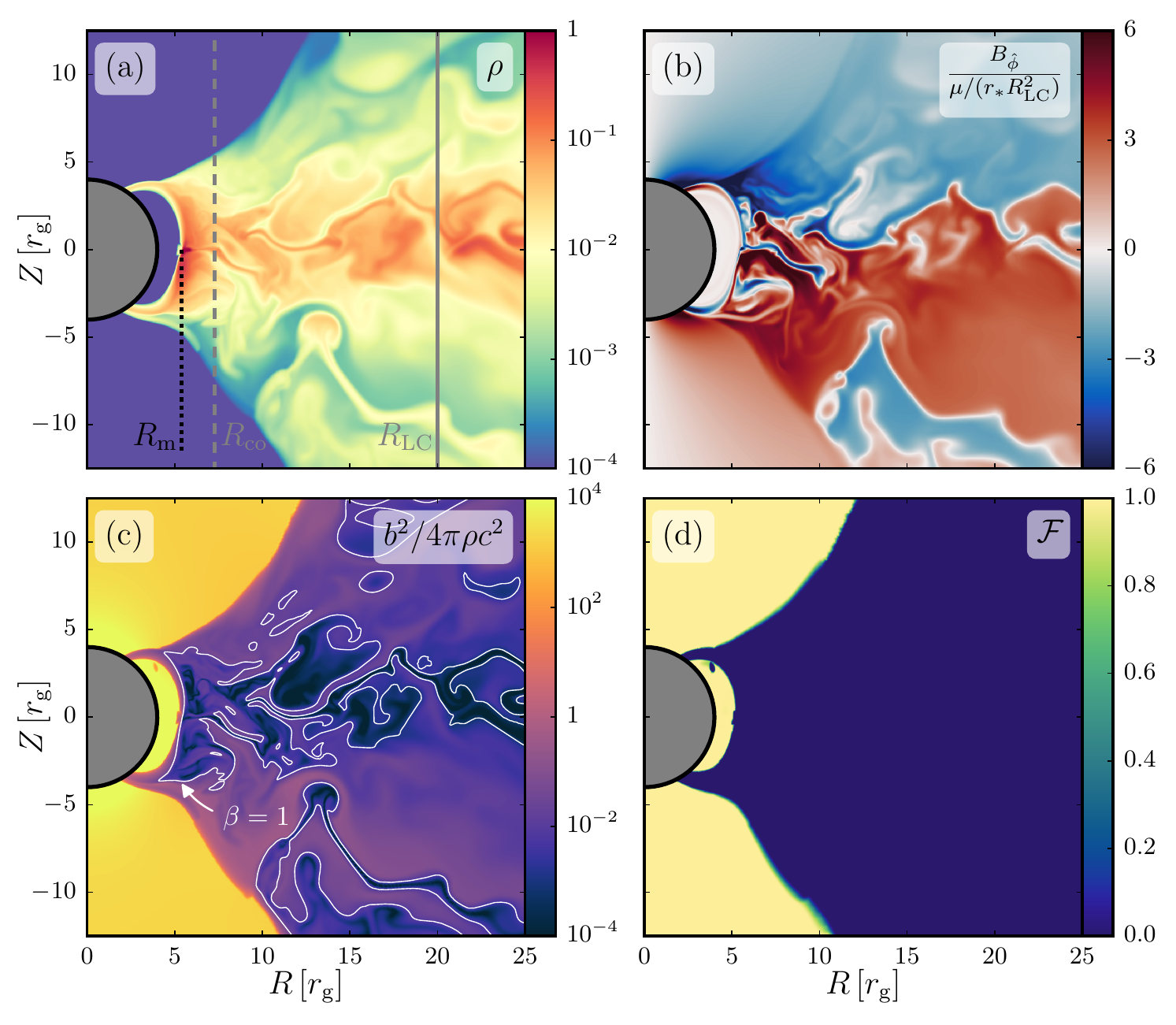}
  \end{center}
  \caption{ \label{fig:variables} 
  Quasi-steady magnetically channeled accretion;
  the $\mu=5$, $\rlc=20\rgu$, 
  anti-parallel magnetic field simulation at $t=16,\!000\rgu/c$. 
  (a) Mass density; vertical lines indicate the corotation (gray dashed), light-cylinder (gray solid),
      and magnetospheric (black dotted) radii.
  (b) Toroidal magnetic field.
  (c) Magnetization $\sigma$, with $\beta=1$ surfaces in white. 
  (d) Passively advected scalar $\ffrac$, the magnetospheric mass-density fraction.} 
\end{figure*}

The isolated-pulsar magnetosphere is established on an Alfv\'{e}n-crossing timescale
\citep[e.g.][]{Contopoulos:1999aa,Spitkovsky:2006aa,Parfrey:2012ab}. 
Rotating open field lines form a pulsar wind that engulfs the torus,
well before the MRI begins to destroy the torus's equilibrium at $t\approx3,\!000\rgc$.
For our simulation with $\mu=5$, $\rlc=20\rgu$, and the anti-parallel orientation of 
stellar dipole and torus magnetic fields, this results in an accretion disk that flows inward against the pulsar
wind, and into the LC. 
The disk's magnetic field is free to reconnect with the star's closed field due to their
anti-parallel orientation, placing the disk's gas on the stellar field lines.
The newly reconnected field line may be added to the pulsar's wind of open magnetic flux, 
centrifugally ejecting the threaded gas.
Otherwise, the stellar field channels the disk material into accretion columns, 
which form at the magnetospheric radius $\rmag$ at which the stellar field truncates the equatorial flow,
as seen in Fig.~\ref{fig:variables}(a). 
Matter enters the columns with low radial velocity and
is gravitationally accelerated to $v_{\hat{r}}\approx-c/3$ at the stellar surface. Our boundary
conditions allow the matter to pass undisturbed through the surface, so no accretion shock forms. 
 
Figure~\ref{fig:variables}(b) shows the smooth regions of `swept back'
toroidal magnetic field\footnote{Measured by the observer normal to spatial
  hypersurfaces: $n_\alpha=(n_t,0_i)$.} connected to 
the stellar poles that form the pulsar wind, spinning down the star.
Matter in the accretion columns along the edge of the
closed zone initially rotates faster than the star and
pulls the stellar field forward, generating $B_{\hat\phi}$ of opposite
sign to that in the wind and delivering a spin-up torque.
The irregularity in $B_{\hat\phi}$ in the disk reflects MHD turbulence down to $\rmag$.

The columns are magnetically dominated ($\sigma>1$) 
at their bases; elsewhere the accretion flow is matter-dominated [Fig.~\ref{fig:variables}(c)]. 
Near the magnetospheric boundary, the accretion flow's pressure is 
largely thermal ($\beta>1$), while $\beta<1$ in the columns,
the mass-loaded wind, and in some of the equatorial regions. 
The magnetospheric regions are strongly magnetically dominated, with $\sigma>10^4$ near the star,
enabled by our method of combining force-free and standard-GRMHD evolution. 
The passive scalar $\ffrac$ is shown in Fig.~\ref{fig:variables}(d): 
$\ffrac\approx0$ indicates plasma from the torus, while 
the $\ffrac\approx1$ region is magnetospheric; 
the absence of plasma with intermediate $\ffrac\sim0.5$ supports this means of 
distinguishing between force-free and MHD regions.

\begin{figure*}
  \begin{center}
    \vspace{-1mm}
    \includegraphics[width=\textwidth, trim = 2.0mm 2.0mm 2.0mm 2.0mm, clip]{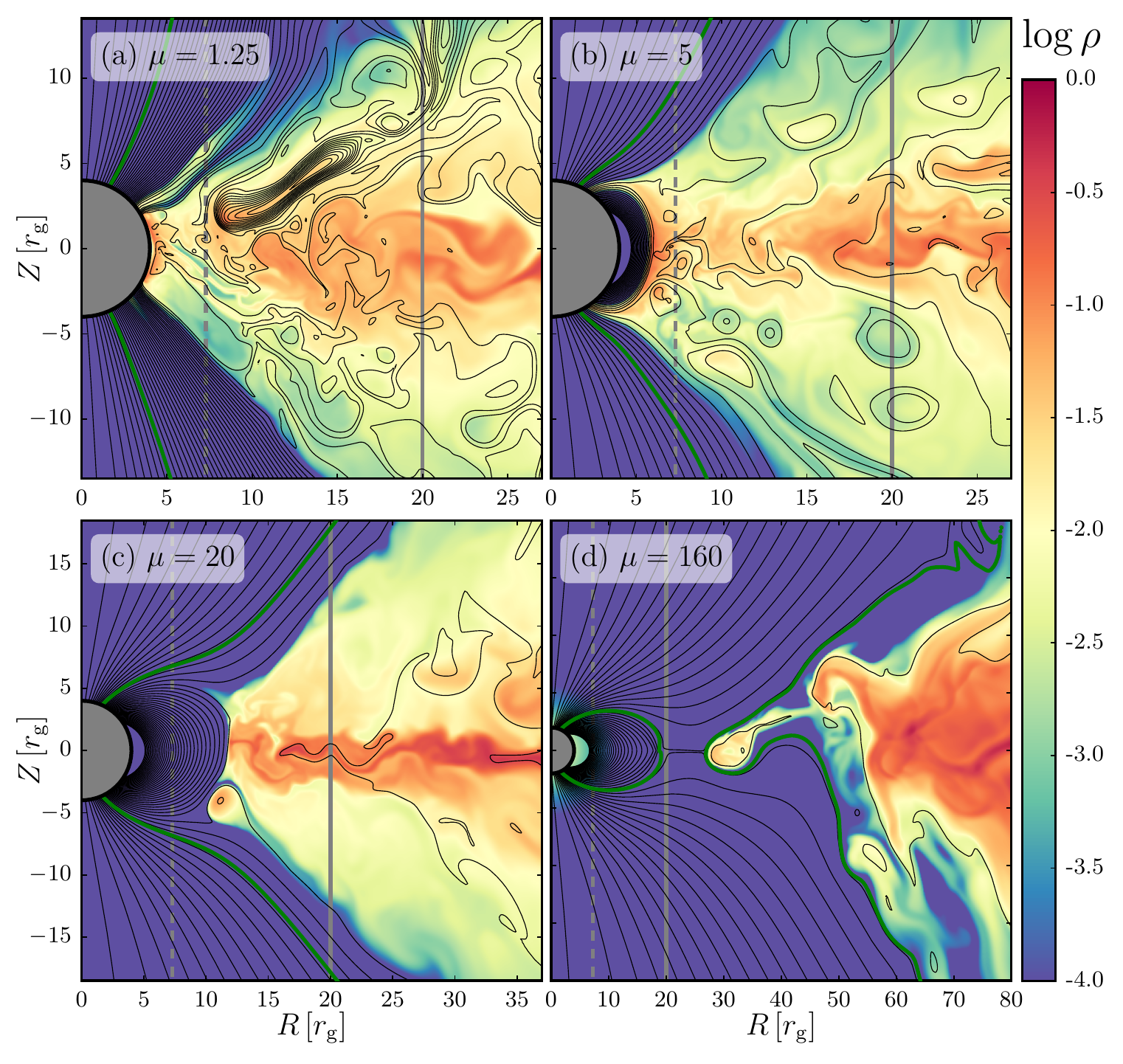}
  \end{center}
  \caption{ \label{fig:fourstates} 
  Four accretion states, determined by the stellar magnetic moment $\mu$ for 
  the same initial torus and stellar spin frequency. 
  Poloidal magnetic field lines are in black, equispaced in magnetic flux, with the thicker
  green line representing the last closed field line for an equivalent isolated pulsar.
  Gray vertical lines indicate $\rco$ and $\rlc$, as in Fig.~\ref{fig:variables}.
  (a) Direct accretion. 
  (b) Magnetically channeled accretion onto the stellar poles. 
  (c) The propeller state. 
  (d) Exclusion of the accretion flow from the light cylinder, shown immediately 
  following the clearing of the inner magnetosphere.} 
\end{figure*}

The location of the magnetospheric radius is determined by the balance between the magnetic
pressure in the closed zone and the hydrodynamic pressure in the inner accretion flow.
By varying the star's magnetic moment $\mu$, while leaving
its rotation rate and the initial gas torus unchanged,\footnote{Equivalent to fixing $\mu$ and 
varying the torus density, and hence the accretion rate.} we can move $\rmag$ in or out, producing distinct states of the combined magnetosphere-disk system.

Figure~\ref{fig:fourstates} illustrates these states (stellar and
torus magnetic fields are initially anti-parallel). 
Whenever the accreting plasma enters the LC there is more open magnetic flux than in the isolated case, leading to a stronger
electromagnetic pulsar wind \citep{Parfrey:2016aa}. This wind is collimated by the 
accretion flow, forming Poynting-flux-dominated relativistic jets.

At the lowest magnetic moment (equivalently, highest accretion rate), almost all of the star's closed field lines are
either opened or crushed, leading to direct accretion near the equator [Figure~\ref{fig:fourstates}(a)].
The flow does not spread across the stellar surface due to our infall boundary conditions.
At larger $\mu$ the star's magnetic field truncates the disk, with accretion proceeding via
magnetically confined columns [Figure~\ref{fig:fourstates}(b)]. 

\begin{figure*}
  \begin{center}
    \includegraphics[width=\textwidth, trim = 2.0mm 2.0mm 2.0mm 2.0mm, clip]{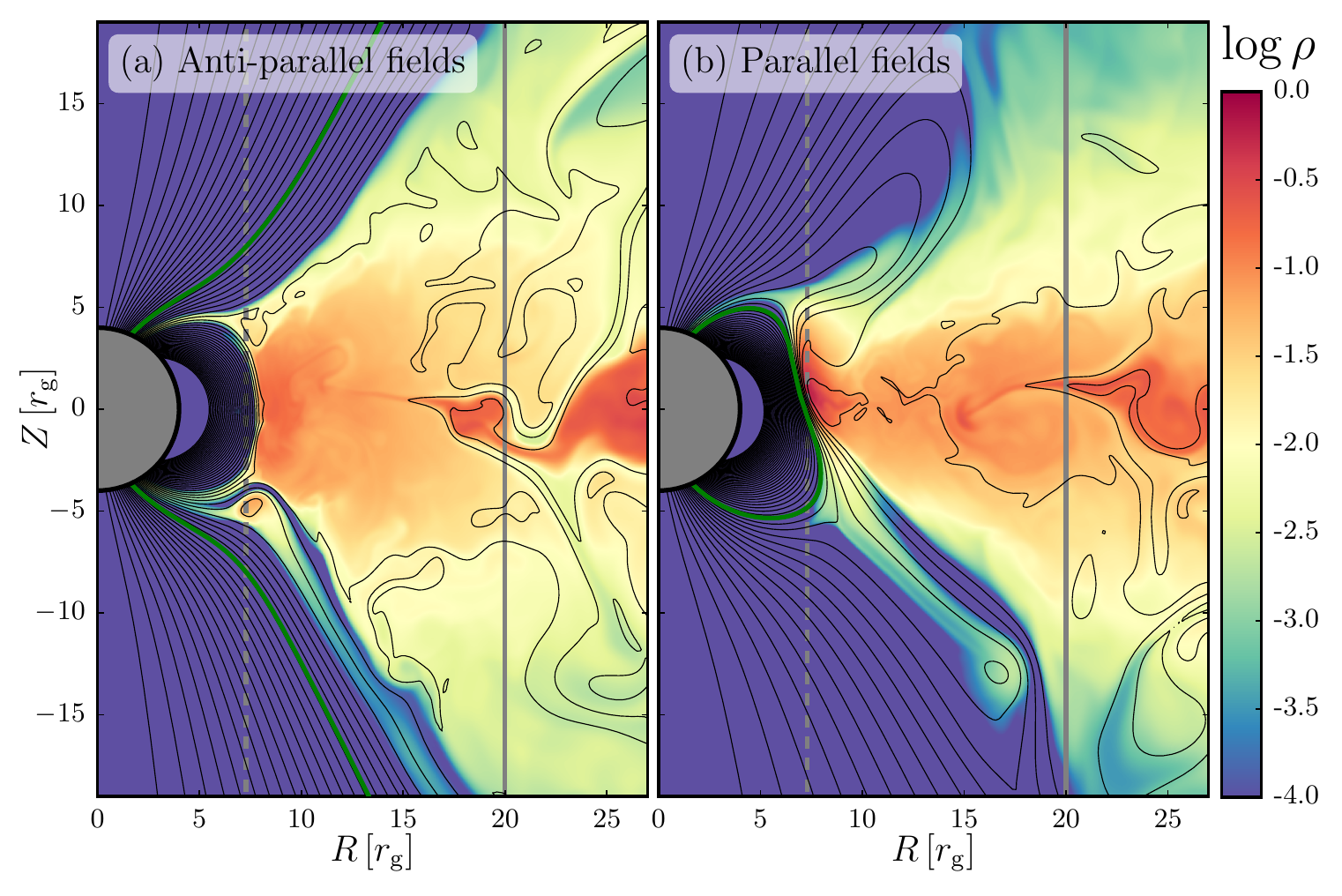}
  \end{center}
  \caption{ \label{fig:fieldOrientation} Effect of relative orientation of the stellar dipole 
    and the initial magnetic field in the torus; $\mu=10$ and $t=21,\!760\rgc$ in both cases. 
    Markings are as in Figure~\ref{fig:fourstates}. 
  (a) Anti-parallel fields: the torus's field lines can directly reconnect with those of 
      the stellar closed zone, facilitating accretion and increasing the open magnetic flux.
  (b) Parallel fields: direct reconnection is not possible, reducing the number of open field lines
      and suppressing accretion.} 
\end{figure*}

At yet higher $\mu$, a state similar to the classical propeller regime is formed [Figure~\ref{fig:fourstates}(c)]
in which the accretion flow is truncated between the corotation radius
$\rco=(\rlc-a\rgu)^{2/3}\rgu^{1/3}$ and $\rlc$. The accretion rate is
very low in this state, with only rare instances of low-density streams reaching the stellar
surface. Matter is centrifugally ejected from near $\rmag$ when reconnection attaches it to the rotating magnetosphere.
An intermediate state, with $\rmag\approx\rco$, is also possible [see Figure~\ref{fig:fieldOrientation}(a)].

At the highest magnetic moment, the pulsar wind is powerful enough
to almost entirely exclude the accretion flow from the LC. 
In the simulation shown in 
Figure~\ref{fig:fourstates}(d) the flow enters the LC four times, each time
only residing for $\approx 900\rgc$, or roughly seven stellar spin periods, before being entirely thrown back
out again. None of the gas reaches the stellar surface.

We find another intermediate state ($\mu = 80$) in which the accretion flow spends most of the 
simulation inside the LC, but is occasionally ejected. When inside the LC the 
flow is strongly churned by interaction with the stellar field.

Our initial conditions, in which the stellar field
lines are deformed around, and do not penetrate, the torus lying beyond $\rlc$, 
allow a clean test of the field-orientation effect 
(Figure~\ref{fig:fieldOrientation}). 
When the direction of the torus's poloidal magnetic loop is reversed, 
the initial field strength $|\vec{B}|$ remains unchanged everywhere in the domain.
Figure~\ref{fig:fieldOrientation}(a) shows that, for anti-parallel fields, reconnection between the stellar
and disk fields permits the opening of originally closed field lines and the formation of
long-lived accretion columns.
This reconnection does not occur in the parallel case, 
and the accreting plasma instead pushes the closed field lines 
inward \citep[e.g.][]{Romanova:2011aa}. 
Rather than opening the stellar field, the accretion flow increases the amount of closed
flux, Figure~\ref{fig:fieldOrientation}(b). Stable columns are not formed, and accretion is
more time-dependent, as plasma diffuses onto the stellar field in the region of large
gradients at $\rmag$.

\begin{figure}
  \begin{center}
    \includegraphics[width=\linewidth, trim = 2.0mm 2.0mm 2.0mm 2.0mm, clip]{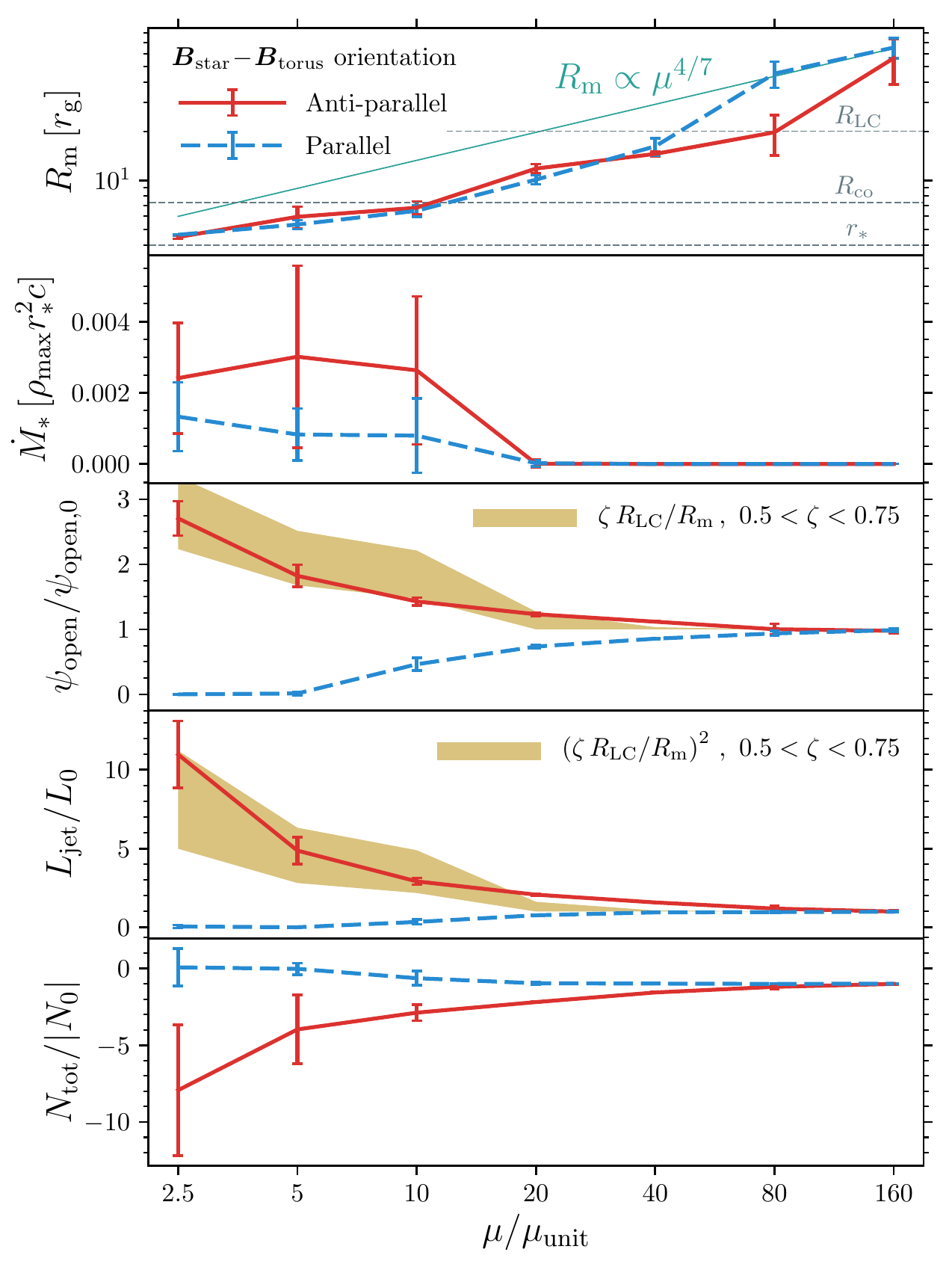}
  \end{center}
  \caption{ \label{fig:muLine} 
  Averaged global quantities as a function of stellar magnetic moment, for \rlc=20\rgu. 
  Vertical bars indicate the standard deviation over the averaging period.
  From top to bottom: magnetospheric radius; mass accretion rate onto the
  stellar surface; magnetic flux open through $r=\rlc$ in the force-free region ($\ffrac>0.5$);
  total torque on the star measured at its surface; electromagnetic jet power.  
  The shaded regions show predictions from a simple flux-opening model.
  Quantities subscripted with `0' indicate their measured isolated-state values; note that 
  $N_0<0$ (i.e., causes spin-down).} 
\end{figure}

\begin{figure}
  \begin{center}
    \includegraphics[width=\linewidth, trim = 2.0mm 2.0mm 2.0mm 2.0mm, clip]{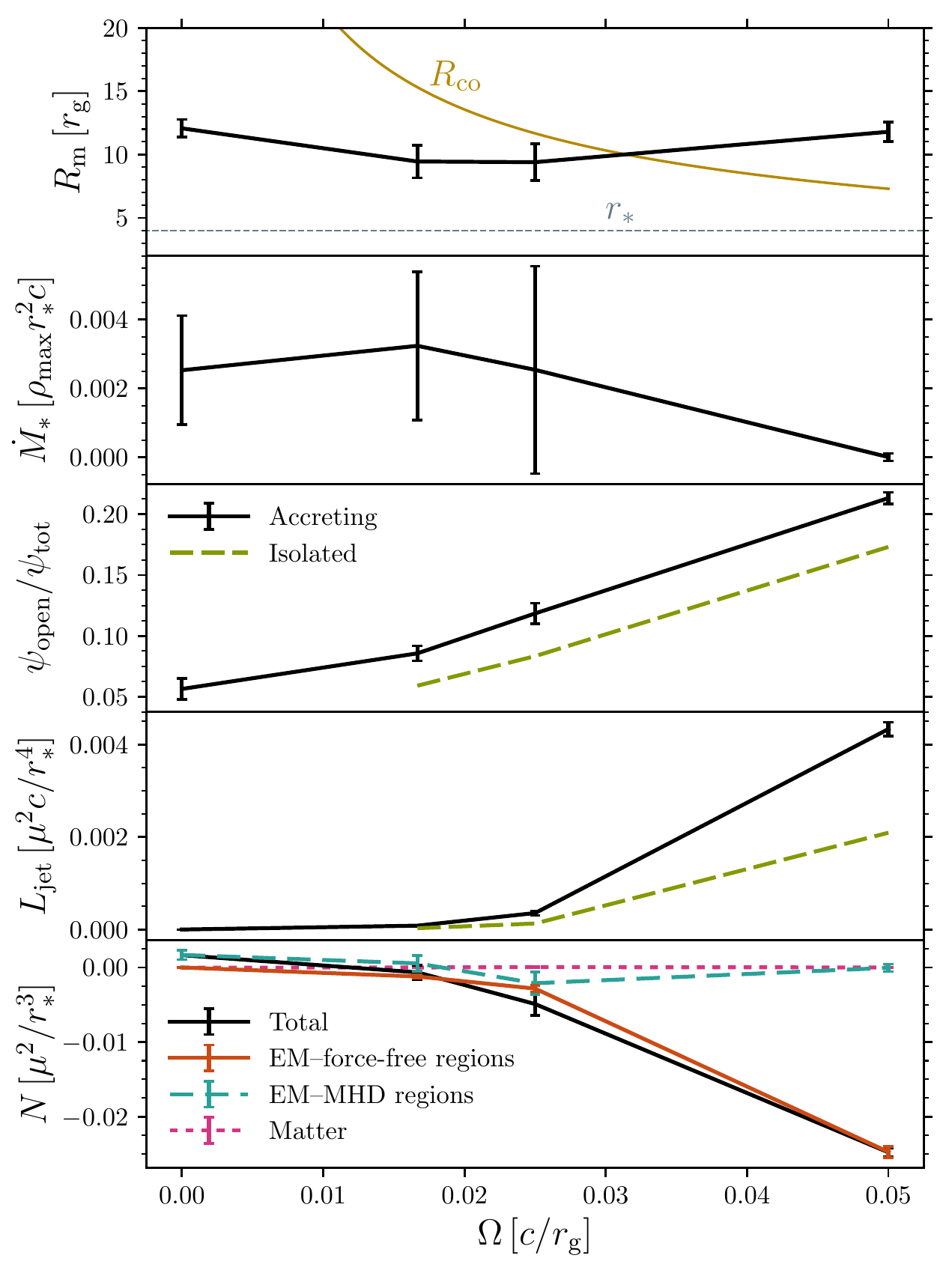}
  \end{center}
  \caption{ \label{fig:omegaLine} 
  Averaged quantities as a function of spin frequency $\Omega=c/\rlc$, for $\mu=20$ and
  the anti-parallel star-torus field orientation. Panels are the same as in Figure~\ref{fig:muLine},
  however, all are now plotted in absolute units; $\psi_{\rm tot}$ is the total magnetic flux 
  through the stellar surface. 
  The dashed green lines indicate the open flux and electromagnetic wind power from an isolated pulsar.} 
\end{figure}

In Figure~\ref{fig:muLine} we show global quantities as a function of stellar magnetic moment $\mu$,
time-averaged between $t=20,\!000$--$50,\!000\rgc$;
vertical lines indicate the standard deviation, and hence the variability.

The magnetospheric radius $\rmag$ increases with increasing $\mu$, and is similar for the 
two field orientations at all $\mu$, as seen instantaneously in Figure~\ref{fig:fieldOrientation}.
The scaling resembles the standard $\rmag\propto\mu^{4/7}$ dependence \citep[e.g.][]{Pringle:1972aa}.

The mass accretion rate is very low or zero when the disk is held beyond corotation ($\mu \geq 20$), 
and is otherwise approximately independent of $\mu$. The rate is 
higher with anti-parallel fields, as stable accretion columns form and gas can 
fall onto the star without diffusing across field lines. The MRI does not create a self-sustaining
dynamo in these axisymmetric simulations, and so the accretion is transient, appearing as a 
long outburst; $\dot{M}$ therefore has a large variability. The other quantities of interest
remain relatively steady. 

The amount of force-free magnetic flux through the $r=\rlc$ surface,
$\psi_{\rm open}$, varies strongly as $\mu$ decreases and $\rmag$ moves inward, with $\psi_{\rm open}$
increasing over its isolated-pulsar value for anti-parallel fields and decreasing to nearly zero in the
parallel case (see also Figure~\ref{fig:fieldOrientation}). 
The relativistic jet power (flux of electromagnetic energy through $r=\rlc$, in the force-free region) 
roughly follows the open flux as $L_{\rm jet}\propto\psi_{\rm open}^2$.
Similarly, the total stellar torque $N_{\rm tot}$ 
becomes increasingly negative (i.e., gives more rapid spin-down) as accretion increases the open
magnetic flux (anti-parallel fields) and 
retreats toward zero torque for the closed-magnetosphere solutions (parallel fields).
Here the open flux, torque, and jet power are
normalized by their values in the absence of accretion; note $L_0=-N_0\Omega=\mu^2\Omega^4/c^3$
\citep{Gruzinov:2005aa}. See \citet{Parfrey:2016aa} for further discussion on using these 
isolated-pulsar values to create dimensionless variables.

The shaded regions in Figure~\ref{fig:muLine} show the range predicted by
a simple model, in which the spin-down torque and jet power are related to the magnetic flux
opened by the accretion flow, estimated via $\psi_{\rm open}=\zeta(\rlc/\rmag)\psi_{\rm open,0}$, 
where $\zeta$ is an efficiency parameter \citep{Parfrey:2016aa,Parfrey:2017aa}.
The jet power is then $L_{\rm jet}=\zeta^2(\rlc/\rmag)^2L_0$. The figure suggests that opening can be
efficient, with $\zeta>0.5$ for anti-parallel fields.

Figure~\ref{fig:omegaLine} illustrates the dependence of the global quantities 
on the stellar spin frequency $\Omega$ ($\mu=20$ in all cases). 
The magnetospheric radius is insensitive to $\Omega$. 
$\dot{M}$ onto the star depends weakly on $\Omega$ when accretion occurs, and is zero
for the propeller system. The open magnetic flux is above its
isolated-pulsar value in all cases, leading to an enhancement in the
power of the electromagnetic wind or jet, which is a strong function
of the stellar spin frequency, $L_{\rm jet}\propto\Omega^{3.6}$.

The bottom panel of Figure~\ref{fig:omegaLine} shows the stellar torque separated into three contributions:
the electromagnetic torque in the force-free ($\ffrac>0.5$) and MHD ($\ffrac<0.5$) regions, and the 
material (or hydrodynamic) torque. The
material torque is negligible in every case; by the time
the accreting gas reaches the surface it is nearly in corotation with the star, having transferred its 
angular momentum to the stellar magnetic field whose stress applies the torque \citep{Romanova:2002aa}.

The electromagnetic torque in the force-free region is always negative (spin-down) or zero, while its average
in the MHD region (i.e., in the accretion columns) is generally positive. 
The simulations cover the range
in total torque from strong spin-down ($\rlc=20\rgu$) to weak spin-up (non-rotating star), with the 
$\rlc=60\rgu$ system in approximate spin equilibrium.

\section{Conclusions}

We have presented a method of combining force-free and standard MHD behavior in a self-consistent simulation in 
curved spacetime,
and used this approach to perform the first relativistic-MHD simulations of accretion onto rotating, magnetized
neutron stars. 
The spin frequencies are in the millisecond range, 
and our initial conditions, in which all of the accreting gas originates from an equilibrium torus beyond
the light cylinder, allow the direct simulation of the transition
between isolated (radio pulsar) and accreting (X-ray pulsar) states. 
The energy-conserving MHD evolution produces a thick accretion flow, whose interaction with the stellar
magnetosphere is relevant to hard-state AMXPs, Z sources, and the pulsing ULXs; recently 
\cite{Takahashi:2017aa} have described a radiation-GRMHD simulation of accretion onto a non-rotating star, 
also aimed at the ULXs.

Our method can be described in covariant language for general spacetimes and coordinates, and may be useful in
various contexts involving both regions with significant hydrodynamic inertia and those that are very light and
magnetically dominated, such as black-hole
jets, gamma-ray bursts, and compact-object mergers. It is straightforward to implement in existing codes, 
and adds only one new scalar variable and its associated advection equation to the MHD system.

By varying the star's magnetic field strength, four states of the star-disk system are produced, determined by the position
of the disk-truncation radius with respect to the problem's other three characteristic locations: the stellar,
corotation, and light-cylinder radii. These states include the three regimes previously explored with
non-relativistic MHD simulations (boundary-layer and magnetically channeled accretion, and the propeller regime),
together with a new state in which the pulsar's electromagnetic wind is powerful enough to prevent the entry of the
accretion flow through the light cylinder \citep[e.g.][]{Shvartsman:1970aa,Burderi:2001aa}. 
The pulsar wind drives trains of shocks and sound waves into the flow, 
possibly producing X-ray emission at large distances.
We also find a new intermediate state, in which the accretion flow is usually inside the light cylinder, 
but is regularly expelled back through it by the star's rotating
magnetic field---this expulsion changes the spin-down rate and potential for low-level stellar accretion,
and may be related to the mode switching of the transitional millisecond pulsars \citep[e.g.][]{Linares:2014aa}.

In our axisymmetric simulations, the direction of the torus's magnetic loop significantly affects the system's
evolution, an effect cleanly isolated by our deformed-dipole initial conditions (see Section.~3). 
When the stellar and torus fields are anti-parallel, the accreting plasma efficiently opens the star's 
closed magnetic field, leading to a pair of powerful relativistic jets that are collimated by the thick accretion flow
\citep{Parfrey:2016aa}. 

This mechanism may have operated in the neutron-star merger GW170817, whose unknown compact remnant produced
a jet of power $\sim 10^{49}$--$10^{50}\ergsec$ \citep{Abbott:2017aa,Margutti:2017aa}. 
The flux-opening jet model predicts 
$L_{\rm jet} \sim 10^{52} \, B_{15}^{6/7} \dot{M}_1^{4/7} \nu_1^2 \ergsec$, where the stellar magnetic field strength,
spin frequency, and accretion rate are normalized as
$B_{15} = B_*/10^{15}\G$, $\nu_1 = \nu_* / 1$~kHz, and $\dot{M}_1 = \dot{M} / 1\, \msun\,\rm{s}^{-1}$.
Similarly, our simulation with the largest ratio of jet power to isolated-pulsar-wind power has 
an average $L_{\rm jet} \approx 11\, L_0$ (see Fig.~\ref{fig:muLine}), giving 
$L_{\rm jet} \approx 6.4\times 10^{50} \, B_{15}^2 \nu_1^4 \ergsec$. These estimates suggest that a rapidly rotating
neutron star with a magnetar-strength field is a viable central engine for the observed short-duration gamma-ray 
burst. More speculatively, an early phase of rapid accretion may have caused the 1.7~s delay between 
gravitational waves and gamma-rays, as the accretion flow may have quenched the jet if it overwhelmed the magnetosphere, occurring if $R_{\rm m} \approx 3.5\, B_{15}^{4/7} \dot{M}_1^{-2/7}\km \ll r_*$
\citep{Parfrey:2016aa}.

In the parallel-fields case, no opening occurs and jets are not produced, at least partly due to the limitations of
axisymmetry, where a poloidal field loop cannot change its direction of circulation. This restriction is lifted
in three-dimensional evolution, which is also necessary for 
creating a self-sustaining MRI dynamo and hence a long-term steady state. 
Though expected to be weaker in 3D, 
the magnitude of this effect on torques and jet powers seen in Figure~\ref{fig:muLine} suggests that
there may be interesting observational consequences to variations in any large-scale net field advected inward
by accretion.

\vspace{3mm}

KP was supported by NASA through Einstein Postdoctoral Fellowship grant number PF5-160142. 
AT was supported by the TAC fellowship and in part under grant no.~NSF~PHY-1125915.
The simulations were performed on the SAVIO cluster provided by the Berkeley Research Computing program at the 
University of California, Berkeley, and on the Stampede cluster of the Texas Advanced Computing Center 
(TACC) of the University of Texas at Austin, via allocation AST150062.

\end{document}